\documentclass[preprint2]{aastex}

\usepackage{graphicx,pstricks}
\usepackage{epsfig}

\slugcomment{Not to appear in Nonlearned J., 45.}

\shorttitle{Energy bursts from deconfinement in high-mass twin stars}
\shortauthors{Alvarez-Castillo et al.}

\begin{document}


\title{Energy bursts from deconfinement in high-mass twin stars}


\author{D.~E.~Alvarez-Castillo\altaffilmark{1,2,a}, M.~Bejger\altaffilmark{b}, D.~Blaschke\altaffilmark{a,1},
P.~Haensel\altaffilmark{b}, L.~Zdunik\altaffilmark{b}
}
\affil{$^a$Institut Fizyki Teoretycznej,
Uniwersytet Wroc{\l}awski,
pl. Maxa Borna 9,
50-204 Wroc{\l}aw,
Poland\\
$^b$N. Copernicus Astronomical Center, Polish Academy of Sciences, Bartycka 18, PL-00-716 Warszawa, Poland}
\altaffiltext{1}{Bogoliubov Laboratory for Theoretical Physics,
Joint Institute for Nuclear Research,
Joliot-Curie Str. 6,
141980 Dubna,
Russia}
\altaffiltext{2}{Instituto de F\'{\i}sica,
Universidad Aut\'onoma de San Luis \\
Potos\'{\i} Av. Manuel Nava 6, San Luis Potos\'{\i}, S.L.P. 78290, \\
M\'exico}

\newpage

\begin{abstract}
We estimate the energy reservoir available in the deconfinement phase transition induced 
collapse of a neutron star to its hybrid star mass twin on the ''third family'' branch, 
using a recent equation of state of dense matter. The available energy corresponding to 
the mass-energy difference between configurations is comparable with energies of the most 
violent astrophysical burst processes. An observational outcome of such a dynamical 
transition might be fast radio bursts, specifically a recent example of a FRB with 
a double-peak structure in its light curve. 
\end{abstract}


\pagebreak

\section{Introduction}

Research on the neutron stars' (NS) equation of state (EoS) is currently a very active area. 
Numerous observations have changed our understanding of the cold, dense nuclear matter in 
compact star interiors. 
Recently, accurate determination of the high mass of about 2 M$_\odot$ for 
PSR J0348+0432 \citep{2013Sci...340..448A} and PSR J1614-2230 \citep{2010Natur.467.1081D} 
pulsars has provided a powerful constraint for the stiffness of the EoS. 
On the other hand, radius measurements are not yet precise enough to rule out some of the 
many alternative EoS models.
Among various estimates, frequency resolved pulse shape analysis for the nearest millisecond pulsar~\citep{2013ApJ...762...96B} supports relatively large radii and offers interesting perspectives for future radius measurements as planned with upcoming 
missions, cf. NICER~\citep{nicer}.

Astrophysical models involving NS can account for the observed powerful gamma ray bursts (GRB, \citealt{1973BAAS....5..322K}) emission, where the typical total energy release is of about $10^{48}$ - $10^{50}$ erg s$^{-1}$~\citep{1978MNRAS.183..359C,1986ApJ...308L..43P,1989Natur.340..126E}. Other energetic phenomena like fast radio bursts (FRB) - millisecond duration radio bursts from seemingly cosmological distances \citep{Lorimer2007}, for which 
a model of a NS collapse to a black hole was proposed \citep{2014A&A...562A.137F}, as well as 
natal pulsar kicks \citep{2005ASPC..328..327P,2001ApJ...549.1111L,2005IJMPA..20.1148K,Berdermann:2006rk,2007ApJ...654..290S} are associated with comparably high amounts of energy.

This Letter considers the energy release during a dynamical NS collapse induced by a deconfinement phase 
transition in the core of a compact star (a {\it corequake}). Unlike previous works 
which estimated the energy reservoir for typical ($\sim 1.4\,M_\odot$) mass stars \citep{1999JPhG...25..971D,2002NuPhS.113..268B,2004A&A...416..991A,Zdunik:2005kh}, we employ a recent EoS model derived in \citet{2015A&A...577A..40B} 
that allows for the formation of a ''third family'' of compact stars near the maximum mass. 
For the first time the corequake scenario is considered in which a {\it high-mass} hadronic NS collapses into a hybrid compact star disconnected from the former by a sequence of unstable configurations (for a recent classification of hybrid stars, see \citealt{Alford:2013aca}). 
The point of instability  can be reached by the hadronic NS in a process of dipole-emission spin-down, 
or accretion-induced spin-up by matter from a companion.

While a direct collapse of a magnetized NS to a black hole is one viable mechanism for the explanation of FRBs \citep{2014A&A...562A.137F}, the NS instability and collapse induced by the deconfinement phase transition of the type discussed here would additionally provide an explanation to a 
double component reported recently for FRB121002 by \citet{Champion:2015} with a light curve peak separation of about 5 ms. 
In this case, the NS would undergo a corequake transition to a meta-stable twin hybrid 
NS configuration, which would generate the first peak, before the ultimate collapse 
to a black hole, generating the second peak, as in \citet{2014A&A...562A.137F} model.

\section{High-mass twin equation of state}

The NS mass twin scenario describes the situation in which two stars have the same 
gravitational mass but different radii due to qualitatively different internal composition: 
one of them is purely hadronic while the other (the {\it twin}) is a hybrid star 
whose core contains quark matter \citep{1968PhRv..172.1325G,1981JPhA...14L.471K,2000NuPhA.677..463S,2000A&A...353L...9G,2015PhRvC..91e5808D}. 
High mass NS twins \citep{2015A&A...577A..40B} can potentially identify a critical 
endpoint in the QCD phase diagram \citep{2015arXiv150305576A, 2013arXiv1310.3803B}. 
At the same time they provide an attractive solution to actual problems of compact star physics discussed 
in \citet{2015arXiv150303834B}: 
the masquerade phenomenon \citep{2005ApJ...629..969A}, the hyperon puzzle \citep{2003astro.ph.12446B}, and the reconfinement problem \citep{2011arXiv1112.6430L,2013A&A...551A..61Z}. 
An assessment of NS twin identification has been recently carried out \citep{2014arXiv1412.8226A,2014arXiv1408.4449A,2014JPhCS.496a2002B}.

NS models presented here are based on the DD2 hadronic EoS with excluded volume correction resulting from the quark substructure of nucleons (protons and neutrons) originating from Pauli blocking effects 
at the level of quark substructure \citep{Ropke:1986qs}.  
At suprasaturation densities this correction is necessary and has the immediate effect of stiffening the EoS. 
High NS masses are reached already at moderate densities but 
large radii ($\simeq 13-15$ km, \citealt{2013arXiv1310.3803B}). 
 At higher densities this EoS undergoes a phase transition 
 to quark matter described by a NJL EoS with multiquark interactions first introduced 
 in \citet{2014EPJA...50..111B}. 
 The coupling strength parameter in the vector channel of the 8-quark interaction ($\eta_4$) 
 determines a sufficient stiffening of the quark matter EoS at high densities
 to preserve the maximum observed star mass of 2~M$_{\odot}$ in this class of models.
\begin{figure}[!t]
\centering
\includegraphics[width=0.6\textwidth]{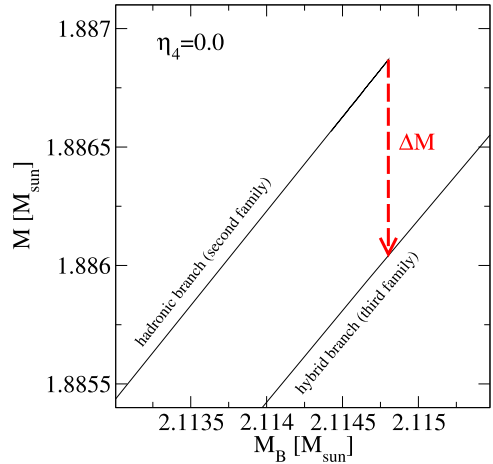}
\caption{Gravitational mass $M$ vs. baryonic mass $M_B$
for the hybrid EoS model with vector coupling parameter $\eta_4=0.0$.
Dynamical NS instability induced by a strong first order phase transition: 
the last stable configuration on the hadronic branch (initial state) is 
connected with the final hybrid star configuration at the same baryonic mass 
with a dashed red arrow line. Length of the line represents the mass difference 
(energy reservoir) occurring in this transition. The results for $\eta_4=$5.0, 10.0 are 
qualitatively and quantitatively similar.}
\label{fig:MvsR}
\end{figure}
\section{Results and discussion}
The EoS of \citet{2015A&A...577A..40B} features the possibility of first order phase transitions 
substantial enough to destabilize the NS. 
The appearance of a small, sufficiently dense quark core makes 
some configurations unstable against radial oscillations (see e.g. \citealt{1986bhwd.book.....S}) 
and is therefore not realized in astrophysical settings. 
On the mass-radius $M(R)$ diagram for non-rotating stars these configurations are characterized by 
$\partial M/\partial \lambda_c < 0$, where $\lambda_c$ is an EoS parameter labelling the configurations 
(e.g., their central density). 
Figure \ref{fig:MvsR} traces the subsequent evolution - a dynamical collapse - 
of an unstable compact star from the point of the maximum mass $M_\mathrm{max,h}$ 
on the hadronic branch to the corresponding one on the hybrid star branch. 
For simplicity we assume that the baryon mass $M_B$ is conserved during the collapse. 
\begin{figure}[h!]
\centering
\includegraphics[width=0.6\textwidth]{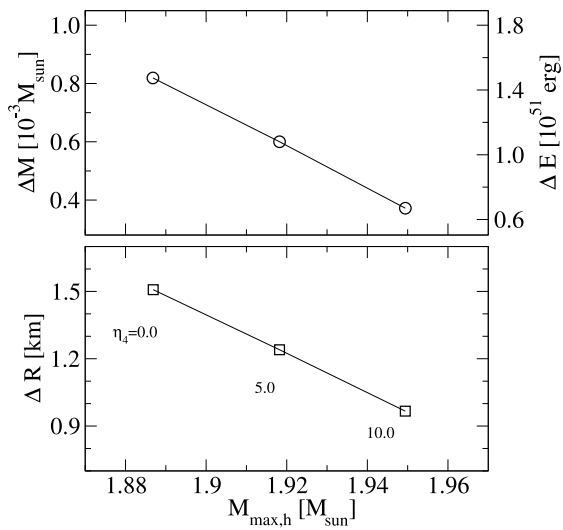}
\caption{Mass difference $\Delta M$ (upper panel) and radius difference $\Delta R$ (lower panel) resulting 
from an instability and subsequent NS collapse induced by a deconfinement phase transition for a set of vector coupling parameters $\eta_4$ of the high-mass twin EoS models. 
An associated energy release $\Delta E$ is indicated on the right side of the upper panel.}
\label{fig:Mdefect}
\end{figure}
Figure~\ref{fig:Mdefect} summarizes the radius change $\Delta R$ in the transition, which 
is between $1$ km and $1.5$ km 
for the coupling constant range considered.
The energy reservoir $\Delta E$ available in the transition, 
defined as the mass-energy difference $\Delta Mc^2$ between the 
initial and final configurations, amounts to roughly $10^{51}$ erg. 
We note that in case of the EoS employed in \citet{Zdunik:2005kh} the energy release 
depends weakly on the rotation rate; rotating configurations and the 
dependence of the energy reservoir on the rotation rate for EoS of \citet{2015A&A...577A..40B} will be studied in a subsequent  article \citep{Bejger:2015}. 

In conclusion, the energy reservoir available in a high-mass NS instability induced by the deconfinement phase transition is comparable with values measured in burst phenomena associated with most energetic astrophysical processes. 
It is clear that the deconfinement phase transition in compact stars might play an important role, perhaps as an engine for GRBs or as an additional mechanism supporting the explosion of core collapse supernovae, with an accompanied characteristic neutrino signals. 
A magnetized NS that is collapsing in a dynamical timescale ($\simeq 1$ ms) to another, compact configuration may also be attractive in the context of FRBs, specifically in view of recent observations of FRB121002 \citep{Champion:2015} with a double peak light curve structure. 
Finally, we note that the NS EoS is a key input for scenarios of cosmic ray generation like  supernova explosions and NS mergers. 
Further studies to expose the details of astrophysical observables in the case of NS instabilities induced 
by dense-matter phase transitions are work in progress \citep{Bejger:2015}.

\acknowledgments
The authors thank Stefan Typel and Sanjin Beni\'c 
for providing the necessary EoS data sets for this work and are 
indebted to Luciano Rezzolla for pointing out the observation 
of the double-peak FRB. 
D.E.A.-C. acknowledges support by the 
Heisenberg-Landau programme,
the Bogoliubov-Infeld programme and by the 
COST Action MP1304 "NewCompStar".
D.B. received support from the Hessian LOEWE initiative through HIC for FAIR.
This work was supported in part by NCN under grant number 
UMO-2014/13/B/ST9/02621.

\clearpage


\begin{thebibliography}{}
\bibitem[Aguilera et al.(2004)]{2004A&A...416..991A} 
Aguilera, D.~N., Blaschke, D., \& Grigorian, H.
\ 2004, \aap, 416, 991 
\bibitem[Alford et al.(2005)]{2005ApJ...629..969A} Alford, M., Braby, M., 
Paris, M., \& Reddy, S.\ 2005, \apj, 629, 969
\bibitem[Alford et al.(2013)]{Alford:2013aca} 
  Alford, M.~G., Han, S., \& Prakash, M.,\ 2013,
  Phys.\ Rev.\ D {88}, 083013
\bibitem[Alvarez-Castillo et al.(2015)]{2014arXiv1408.4449A} 
  Alvarez-Castillo, D. E., Ayriyan, A., Blaschke, D., \& Grigorian, H.\ 2015,
  arXiv:1506.07755
\bibitem[Alvarez-Castillo \& Blaschke(2015)]{2015arXiv150305576A} 
  Alvarez-Castillo, D.~E., \& Blaschke, D.\ 2015, PoS CPOD 2014 (2014) 045 
\bibitem[Antoniadis et al.(2013)]{2013Sci...340..448A} Antoniadis, J., 
Freire, P.~C.~C., Wex, N., et al.\ 2013, Science, 340, 448
\bibitem[Arzoumanian et al.(2014)]{nicer} Arzoumanian, Z., 
Gendreau, K.~C., Baker, C.~L., et al.\ 2014, \procspie, 9144, 914420
\bibitem[Ayriyan et al.(2014)]{2014arXiv1412.8226A} Ayriyan, A., 
Alvarez-Castillo, D.~E., Blaschke, D., Grigorian, H., 
\& Sokolowski, M.\ 2014, arXiv:1412.8226
 \bibitem[Baldo et al.(2003)]{2003astro.ph.12446B} 
 Baldo, M., Burgio, G.~F., \& Schulze, H.~-J.\ 2003,  
 {\it Superdense QCD Matter and Compact Stars}, 
 D. Blaschke and D. Sedrakian (Eds.),
 Springer, Heidelberg, 2006, p.113
\bibitem[Bejger et al.(2015)]{Bejger:2015}
Bejger, M., et al.,\ 2015, in preparation
\bibitem[Beni{\'c}(2014)]{2014EPJA...50..111B} 
Beni{\'c}, S.\ 2014,  Eur. Phys. J. A, 50, 111 
 \bibitem[Beni{\'c} et al.(2015)]{2015A&A...577A..40B} 
 Beni{\'c}, S., Blaschke, D., Alvarez-Castillo, D.~E., Fischer, T., \& Typel, S.\ 2015, \aap, 577, A40
\bibitem[Berdermann et al.(2005)]{Berdermann:2006rk} 
  Berdermann,~J., Blaschke, D., Grigorian, H., \& Voskresensky, D.~N.\ 2006, 
  Prog.\ Part.\ Nucl.\ Phys.\  {57}, 334
\bibitem[Berezhiani et al.(2002)]{2002NuPhS.113..268B} 
Berezhiani, Z., Bombaci, I., Drago, A., Frontera, F., \& Lavagno, A.
\ 2002, Nucl. Phys. B Proc. Suppl., 113, 268 
 \bibitem[Blaschke et al.(2013)]{2013arXiv1310.3803B} 
  Blaschke, D., Alvarez-Castillo, D.~E., \& Benic, S.\ 2013, PoS CPOD {2013} (2013), 063
\bibitem[Blaschke et al.(2014)]{2014JPhCS.496a2002B} 
Blaschke, D.~B., Grigorian, H.~A., Alvarez-Castillo, D.~E., \& Ayriyan, A.~S.\ 2014, 
J. Phys. Conf. Ser., 496, 012002
 \bibitem[Blaschke \& Alvarez-Castillo(2015)]{2015arXiv150303834B} 
 Blaschke, D., \& Alvarez-Castillo, D.~E.\ 2015, arXiv:1503.03834 
\bibitem[Bogdanov(2013)]{2013ApJ...762...96B} 
 Bogdanov, S.\ 2013, \apj, 762, 96 
\bibitem[Cavallo \& Rees(1978)]{1978MNRAS.183..359C} 
Cavallo, G., \& Rees, M.\ 1978, \mnras, 183, 359
\bibitem[Champion(2015)]{Champion:2015}
Champion, D.,\ 2015, talk at seventh Bonn Workshop on "Formation and Evolution of Neutron Stars",
May 18, 2015
\bibitem[Demorest et al.(2010)]{2010Natur.467.1081D} Demorest, P.~B., 
Pennucci, T., Ransom, S.~M., Roberts, M.~S.~E., \& Hessels, J.~W.~T.\ 2010, \nat, 467, 1081
\bibitem[Dexheimer et al.(2015)]{2015PhRvC..91e5808D} Dexheimer, V., 
Negreiros, R., \& Schramm, S.\ 2015, \prc, 91, 055808
\bibitem[Drago \& Tambini(1999)]{1999JPhG...25..971D} 
Drago, A., \& Tambini, U.
\ 1999, J. Phys. G, 25, 971 
\bibitem[Eichler et al.(1989)]{1989Natur.340..126E} Eichler, D., Livio, M., 
Piran, T., \& Schramm, D.	~N.
\ 1989, \nat, 340, 126 
\bibitem[Falcke \& Rezzolla(2014)]{2014A&A...562A.137F} Falcke, H., \& Rezzolla, L.
\ 2014, \aap, 562, A137
\bibitem[Gerlach(1968)]{1968PhRv..172.1325G} 
Gerlach, U.~H.\ 1968, Phys. Rev., 172, 1325
\bibitem[Glendenning \& Kettner(2000)]{2000A&A...353L...9G} 
Glendenning, N.~K., \& Kettner, C.\ 2000, \aap, 353, L9 
\bibitem[K\"ampfer(1981)]{1981JPhA...14L.471K} 
K\"ampfer, B.\ 1981, J. Phys. A, 14, L47
\bibitem[Klebesadel et al.(1973)]{1973BAAS....5..322K} Klebesadel, R.~W., 
Strong, I.~B., \& Olson, R.~A.\ 1973, \baas, 5, 322
\bibitem[Kusenko(2005)]{2005IJMPA..20.1148K} 
Kusenko A.
\ 2005, Int. J. Mod. Phys. A, 20, 1148 \ 2001, \apj, 549, 1111
\bibitem[Lai et al.(2001)]{2001ApJ...549.1111L} Lai, D., Chernoff, D.~F., 
\& Cordes, J.~M.
\ 2001, \apj, 549, 1111 
\bibitem[Lastowiecki et al.(2011)]{2011arXiv1112.6430L} Lastowiecki, R., 
Blaschke, D., Grigorian, H., \& Typel, S.,\ 2012, 
Acta Phys.\ Polon.\ Supp.\   5, 535
\bibitem[Lorimer et al.(2007)]{Lorimer2007} 
Lorimer, D.~R., Bailes, M., McLaughlin, M.~A., Narkevic, D.~J., 
\& Crawford, F.\ 2007, Science, 318, 777
\bibitem[Paczynski(1986)]{1986ApJ...308L..43P} Paczynski, B.,
\ 1986, \apjl, 308, L43 
\bibitem[Podsiadlowski et al.(2005)]{2005ASPC..328..327P} Podsiadlowski, 
P., Pfahl, E., \& Rappaport, S.
\ 2005, Binary Radio Pulsars, 328, 327
\bibitem[R\"opke et al.(1986)]{Ropke:1986qs} 
  R\"opke, G., Blaschke,  D., \& Schulz, H.,\ 1986,
  Phys.\ Rev.\ D  34, 3499
\bibitem[Schertler et al.(2000)]{2000NuPhA.677..463S} 
Schertler, K., Greiner, C., Schaffner-Bielich, J., \& Thoma, M.~H.\ 2000, 
Nucl. Phys. A, 677, 463 
\bibitem[Shapiro \& Teukolsky(1986)]{1986bhwd.book.....S} 
Shapiro, S.~L., Teukolsky, S.~A., 
{\it Black Holes, White Dwarfs and Neutron Stars: The Physics of Compact Objects},
\ 1986, Wiley, New York
 \bibitem[Stasielak et al.(2007)]{2007ApJ...654..290S} 
 Stasielak, J., Biermann, P.~L., \& Kusenko, A.\ 2007, \apj, 654, 290 
\bibitem[Zdunik et al.(2008)]{Zdunik:2005kh} 
 Zdunik, J.~L., Bejger, M., Haensel, P., \& Gourgoulhon, E.,
\ 2008, \aap, {479}, 515
\bibitem[Zdunik \& Haensel(2013)]{2013A&A...551A..61Z} 
Zdunik, J.~L., \& Haensel, P.\ 2013, \aap, 551, A61 
\end{thebibliography}
\end{document}